\documentclass{elsart}
\usepackage{epsfig}
    \newcommand{\ba}{\begin{eqnarray}}
    \newcommand{\ea}{\end{eqnarray}}
    \newcommand{\be}{\begin{equation}}
    \newcommand{\ee}{\end{equation}}

    \newcommand{\AmS}{{\protect\the\textfont2%
  A\kern-.1667em\lower.5ex\hbox{M}\kern-.125emS}}

\newcommand{\bn}{{\bf n}}
\newcommand {\bk} {{\mathbf k}}
\newcommand {\bfr} {{\mathbf r}}
\newcommand {\bp} {{\mathbf p}}

\newcommand{\calH}{{\mathcal H}}

\newcommand{\calM}{{\mathcal M}}
\newcommand{\calZ}{{\mathcal Z}}
\newcommand{\calY}{{\mathcal Y}}
\newcommand{\calW}{{\mathcal W}}

\begin{document}
\runauthor{PKU}
\begin{frontmatter}

\title{Two Particle States in an Asymmetric Box}
\author[PKU]{Xin Li}
\author[PKU]{and Chuan Liu}

\address[PKU]{Department of Physics\\
          Peking University\\
                  Beijing, 100871, P.~R.~China}
                \thanks{This work is supported by the National Natural
 Science Foundation (NFS) of China under grant
 No. 90103006, No. 10235040 and supported by the
 Trans-century fund from Chinese
 Ministry of Education.}

\begin{abstract}
The exact two-particle energy eigenstates in an asymmetric
rectangular box with periodic boundary conditions in all three
directions are studied. Their relation with the elastic scattering
phases of the two particles in the continuum are obtained. These
results can be viewed as a generalization of the corresponding
formulae in a cubic box obtained by L\"uscher before.  In
particular, the $s$-wave scattering length is related to the
energy shift in the finite box. Possible applications of these
formulae are also discussed.
\end{abstract}
\end{frontmatter}


\section{Introduction}

 In a series of papers, L\"uscher obtained results
 \cite{luscher86:finiteb,luscher90:finite,luscher91:finitea,luscher91:finiteb} which relate
 the energy of a two-particle state in a cubic box (a
 torus) with the elastic scattering phases of the two
 particles in the continuum. This formula, now known as
 L\"uscher's formula, has been utilized in a number of
 applications, e.g. linear sigma model in the broken phase \cite{Zimmermann94},
 and also in quenched QCD ~\cite{gupta93:scat,fukugita95:scat,jlqcd99:scat,%
 JLQCD02:pipi_length,chuan02:pipiI2,juge03:pipi_length,CPPACS03:pipi_phase,ishizuka03:pipi_length}.
 Due to limited numerical computational power,
 the $s$-wave scattering length, which is related to the scattering
 phase shift at vanishing relative three momentum,
 is mostly studied in hadron scattering using quenched approximation.
 CP-PACS collaboration calculated the scattering
 phases at non-zero momenta in pion-pion $s$-wave scattering in the $I=2$
 channel~\cite{CPPACS03:pipi_phase} using quenched Wilson
 fermions and recently also in two flavor
 full QCD \cite{CPPACS03:pipi_phase_unquench}.
 In typical lattice QCD calculations,
 if one would like to probe for physical
 information concerning two-particle states with non-zero relative three momentum,
 large lattices have to be used which usually requires
 enormous amount of computing resource.
 One of the reasons for this is the following.
 In a cubic box, the three momenta of a single particle
 are quantized according to:
 $\bk=(2\pi/L)\bn\equiv(2\pi/L)(n_1,n_2,n_3)$,
 with $\bn\in Z^3$.
 \footnote{That is, $n_1$, $n_2$ and $n_3$ are integers.}
 In order to control lattice artifacts due to these
 non-zero momentum modes, one needs to have large values of $L$.
 One disadvantage of the cubic box
 is that the energy of a free particle with lowest non-zero momentum
 is degenerate. This means that the second lowest
 energy level of the particle
 with non-vanishing momentum corresponds to $\bn=(1,1,0)$.
 If one would like to measure these states
 on the lattice, even larger values of $L$ should be used.
 One way to remedy this is to use
 a three dimensional box whose shape is not cubic.
 If we use a {\em generic} rectangular box of
 size $(\eta_1L)\times(\eta_2L)\times L$ with $\eta_1$ and
 $\eta_2$ other than unity, we
 would have three different low-lying one-particle energy with
 non-zero momenta corresponding to $\bn=(1,0,0)$, $(0,1,0)$
 and $(0,0,1)$, respectively.
 This scenario is useful since it presents more available low
 momentum modes for a given lattice size,
 which is important in the study of hadron-hadron scattering
 phase shift. Similar situation also occurs in the study of
 $K$ to $\pi\pi$ matrix element (see Ref. \cite{ishizuka02:Kpipi_review} for
 a review and references therein). There, one also needs to study
 two-particle states with non-vanishing relative three-momentum.
 Again, a cubic box yields too few available low-lying
 non-vanishing momenta and large value of $L$ is needed to
 reach the physical interesting kinematic region.
 In these cases,
 one could try an asymmetric rectangular box with only
 one side being large while the other sides moderate.
 In an asymmetric rectangular box, the original formulae
 due to L\"uscher, which give the relation between the
 energy eigenvalues in the finite box and the continuum scattering
 phases, have to be modified accordingly.
 The purpose of this letter is to derive the equivalents of
 L\"uscher's formulae in the case of a generic
 rectangular (not necessarily cubic) box.

 We consider two-particle states in a box of size
 $(\eta_1L)\times (\eta_2L)\times L$ with periodic boundary
 conditions. For definiteness, we take
 $\eta_1\geq 1$,$\eta_2\geq 1$, which amounts to denoting the
 the length of the smallest side of the rectangular box as $L$.
 The following derivation depends heavily on the previous
 results obtained in Ref.~\cite{luscher91:finitea}.
 We will take over similar assumptions as in Ref.~\cite{luscher91:finitea}.
 In particular, the relation between
 the energy eigenvalues and the scattering
 phases derived in the non-relativistic quantum mechanical model can be carried
 over to the case of relativistic, massive field theory under these assumptions,
 the same way as in the case of cubic box which was
 discussed in detail in Ref.~\cite{luscher91:finitea}.
 For the quantum mechanical model,
 we assume that the range of the interaction, denoted by $R$,
 of the two-particle system is such that $R<L/2$.

 The modifications which have to be implemented,
 as compared with Ref.~\cite{luscher91:finitea},
 are mainly concerned with different symmetries of the box.
 In a cubic box, the representations
 of the rotational group are decomposed into irreducible
 representations of the cubic group. In a generic asymmetric box,
 the symmetry of the system is reduced. In the case
 of $\eta_1=\eta_2\neq 1$, the basic group becomes $D_4$;
 if $\eta_1\neq \eta_2 \neq 1$, the symmetry is further reduced to $D_2$,
 modulo parity operation.
 Therefore, the final expression relating the energy eigenvalues of
 the system and the scattering phases will be different.

 \section{Energy eigenstates and singular periodic solutions of Helmholtz equation}
 \label{sec:SPS}

 As discussed in Ref.~\cite{luscher91:finitea}, the energy
 eigenstates in a box is intimately related to the singular
 periodic solutions of the Helmholtz equation:
 \be
 (\nabla^2+k^2)\psi(\bfr)=0 \;.
 \ee
 These solutions are periodic:
 $\psi(\bfr+\hat{\bn}L)=\psi(\bfr)$
 and are bounded by certain powers of $r=|\bfr|$ near $r=0$.
 The momentum modes in the rectangular box are quantized as:
 $\bk=(2\pi/L)\tilde{\bn}$.
 Here we introduce the notations:
 $\tilde{\bn}\equiv (n_1/\eta_1,n_2/\eta_2,n_3)$ and
 $\hat{\bn}\equiv (n_1\eta_1,\eta_2n_2,n_3)$ with
 $\bn=(n_1,n_2,n_3)\in Z^3$.
 For regular values of $k$,
 \footnote{This means that $|k|\neq (2\pi/L)|\tilde{\bn}|$ for
 any $\bn\in Z^3$.}%
 the singular periodic solutions
 of Helmholtz equation can be obtained from the Green's function:
 \be
 G(\bfr;k^2)={1\over\eta_1\eta_2L^3}\sum_\bp
 {e^{i\bp\cdot\bfr}\over \bp^2-k^2}\;.
 \ee
 We define: $\calY_{lm}(\bfr)\equiv r^lY_{lm}(\Omega_\bfr)$,
 where $\Omega_\bfr$\ represents the solid angle
 parameters $(\theta,\phi)$ of $\bfr$ in spherical coordinates;
 $Y_{lm}$ are the usual spherical harmonic functions.
 It is well-known that $\calY_{lm}(\bfr)$ consist
 of all linear independent, homogeneous functions in $(x,y,z)$ of
 degree $l$ that transform irreducibly under the rotational group.
 We then define:
 \be
 G_{lm}(\bfr;k^2)=\calY_{lm}(\nabla)G(\bfr;k^2)\;,
 \ee
 which form a complete, linear independent set of
 functions of singular periodic solutions to
 Helmholtz equation.
 The functions $G_{lm}(\bfr;k^2)$ may be expanded into
 spherical harmonics:
 \be
 G_{lm}(\bfr;k^2)={(-)^lk^{l+1}\over 4\pi}
 \left[Y_{lm}(\Omega_\bfr)n_l(kr)
 +\sum_{l'm'}\calM_{lm;l'm'}Y_{l'm'}(\Omega_\bfr)
 j_{l'}(kr)\right]\;.
 \ee
 Here, $j_l$ and $n_l$ are the usual spherical Bessel functions
 and the matrix $\calM_{lm;l'm'}$ is related to
 the {\em modified} zeta function via:
 \ba
 \label{eq:calM-zeta}
 \calM_{lm;js}&=&
 \sum_{l'm'}{(-)^si^{j-l}\calZ_{l'm'}(1,q^2;\eta_1,\eta_2)\over
 \eta_1\eta_2\pi^{3/2}q^{l'+1}}
 \sqrt{(2l+1)(2l'+1)(2j+1)}
 \nonumber \\
 &\times &
 \left(\begin{array}{ccc}
 l & l' & j \\
 0 & 0  & 0 \end{array}
 \right)
 \left(\begin{array}{ccc}
 l & l' & j \\
 m & m' & -s \end{array}
 \right)\;.
 \ea
 In this formula, the Wigner $3j$-symbols can be
 related to the Clebcsh-Gordan coefficients in the
 usual way.
 For a given angular momentum cutoff $\Lambda$,
 the quantity $\calM_{lm;l'm'}$ can be viewed as the
 matrix element of a linear operator $\hat{M}$ in
 a vector space $\calH_\Lambda$, which is spanned
 by all harmonic polynomials of degree $l\leq\Lambda$.
 The modified zeta function is formally defined by:
 \be
 \label{eq:zeta_def}
 \calZ_{lm}(s,q^2;\eta_1,\eta_2)=
 \sum_{\bn} {\calY_{lm}(\tilde{\bn})
 \over (\tilde{\bn}^2-q^2)^s}\;.
 \ee
 According to this definition,
 the modified zeta function at the right-hand side of Eq.~(\ref{eq:calM-zeta})
 is formally divergent and needs to be analytically continued.
 Following similar discussions as in Ref.~\cite{luscher91:finitea},
 one could obtain a finite expression for the modified zeta
 function which is suitable for numerical evaluation.
 It is also obvious from the symmetry of $D_4$ or $D_2$
 that, for $l\leq 4$, the only non-vanishing zeta functions
 at $s=1$ are: $\calZ_{00}$, $\calZ_{20}$, $\calZ_{2\pm 2}$,
 $\calZ_{40}$, $\calZ_{4\pm 2}$ and $\calZ_{4\pm 4}$.
 One easily verifies that, if $\eta_1=\eta_2=1$, all of the above
 definitions and formulae reduce to the
 those obtained in Ref.~\cite{luscher91:finitea}.

 The energy eigenstates of the two-particle system may be expanded
 in terms of singular periodic solutions
 of Helmholtz equation. This solution in the region where
 the interaction is vanishing can be expressed in terms of
 ordinary spherical Bessel functions, which is related to
 the scattering phases in the usual way.
 For the two-particle eigenstate in the symmetry sector $\Gamma$
 in a box of particular symmetry (either $D_4$ or $D_2$),
 the energy eigenvalue, $E=k^2/2\mu$ with $\mu$ being
 the reduced mass of the two particles, is determined by:
 \be
 \label{eq:main_result}
 \det [e^{2i\delta}-\hat{U}(\Gamma)]=0 \;,\;\;
 \hat{U}(\Gamma)=(\hat{M}(\Gamma)+i)/(\hat{M}(\Gamma)-i)\;.
 \ee
 Here $\Gamma$ denotes a particular representation of the group
 $D_4$ or $D_2$. $\hat{M}(\Gamma)$ represents a linear operator in
 the vector space $\calH_\Lambda(\Gamma)$.
 This vector space is spanned by all complex vectors
 whose components are $v_{ln}$, with $l\leq\Lambda$,
 and $n$ runs from $1$ to the number of occurrence of
 $\Gamma$ in the decomposition of representation with angular
 momentum $l$, see Ref.~\cite{luscher91:finitea} for details.
 To write out more explicit formulae,
 one therefore has to consider decompositions
 of the rotational group representations under
 appropriate symmetries.

 \section{Symmetry of an asymmetric box}
 \label{sec:symmetry}

 As mentioned in the beginning of this letter,
 modifications have to be made since the symmetry
 of an asymmetric box is different from that of a cubic one.
 We first describe the case $\eta_1=\eta_2$. The
 symmetry group is $D_4$, which has $4$ one-dimensional
 representations: $A_1$, $A_2$, $B_1$, $B_2$ and a
 two-dimensional irreducible representation $E$.
 The representations of the rotational group
 are decomposed according to:
 \be
 \label{eq:decomposition_D4}
 {\mathbf 0}=A^+_1\;,\;
 {\mathbf 1}=A^-_2+E^-\;,\;
 {\mathbf 2}=A^+_1+B^+_1+B^+_2+E^+\;,
 \ee
 As is seen, in the $A^+_1$ sector, up to $l\leq 2$
 both $s$-wave and $d$-wave contribute. This corresponds
 to {\em two} linearly independent, homogeneous polynomials
 with degrees not more than $2$, which are invariant under $D_4$.
 These two polynomials can be identified as $\calY_{00}$ and
 $\calY_{20}\propto (x^2+y^2-2z^2)$.
 Therefore, we can write out the reduced matrix element
 $\calM(A^+_1)_{ll'}=m_{ll'}$ in this sector:
 \ba
 \label{eq:matrix_D4}
 m_{00}&=&\calW_{00}\;,\;
 m_{20}=m_{02}=-\calW_{20}\;,\nonumber \\
 m_{22}&=&\calW_{00}+{2\sqrt{5}\over 7}\calW_{20}
 +{6\over 7}\calW_{40}\;,
 \ea
 where we have introduced the notation:
 \be
 \label{eq:W_def}
 \calW_{lm}(1,q^2;\eta_1,\eta_2)\equiv
 {\calZ_{lm}(1,q^2;\eta_1,\eta_2)
 \over \pi^{3/2}\eta_1\eta_2q^{l+1}}\;.
 \ee
 We find that, in the case of $D_4$ symmetry,
 Eq.~(\ref{eq:main_result}) becomes:
 \be
 \label{eq:result_D4}
 \left(1+{m^2_{02}\over m_{00}}
 {\tan\delta_2\over 1-m_{22}\tan\delta_2}
 \right)\tan\delta_0={1\over m_{00}}\;.
 \ee
 Similar formula also appears in the case of cubic
 box except that the mixing with $s$-wave
 comes in at $l=4$, not at $l=2$.

 For the case $\eta_1\neq\eta_2$, the symmetry
 group becomes $D_2$ which has only $4$ one-dimensional
 representations: $A$, $B_1$, $B_2$ and $B_3$.
 The decomposition~(\ref{eq:decomposition_D4}) is replaced by:
 \be
 \label{eq:decomposition_D2}
 {\mathbf 0}=A^+\;,\;
 {\mathbf 1}=B^-_1+B^-_2+B^-_3\;,\;
 {\mathbf 2}=A^++A^++B^+_1+B^+_2+B^+_3\;.
 \ee
 So, up to $l\leq 2$, $A^+$ occurs {\em three} times:
 once in $l=0$ and twice in $l=2$. The corresponding
 basis polynomials can be taken as: $\calY_{00}$,
 $\calY_{20}$ and $(\calY_{22}+\calY_{2-2})/\sqrt{2}$.
 If we denote the above three states as: $0$, $2$ and $\bar{2}$,
 the reduced matrix $\hat{M}(A^+)$ is three-dimensional
 with matrix elements $m_{00}$,
 $m_{02}=m_{20}$ and $m_{22}$
 given in Eq.~(\ref{eq:matrix_D4}) and the rest are given by:
 \ba
 \label{eq:matrix_D2}
 m_{0\bar{2}}&=&m_{\bar{2}0}=-{1\over\sqrt{2}}(\calW_{22}+\calW_{2-2})\;,
 \nonumber \\
 m_{2\bar{2}}&=&m_{\bar{2}2}=-{\sqrt{10}\over 7}
 (\calW_{22}+\calW_{2-2})+{\sqrt{30}\over 14}(\calW_{42}+\calW_{4-2})\;,
 \nonumber \\
 m_{\bar{2}\bar{2}}&=& \calW_{00}-{2\sqrt{5}\over 7}\calW_{20}
 +{1\over 7}\calW_{40}+\sqrt{{5\over 14}}(\calW_{44}+\calW_{4-4})\;.
 \ea
 Similar to Eq.~(\ref{eq:result_D4}), the relation between
 the energy eigenvalue and the scattering phases now reads:
 \ba
 \label{eq:result_D2}
 &&\left({1\over m_{00}}\cot\delta_0-1\right)
 \left|\begin{array}{cc}
 1-m_{22}\tan\delta_2 &   m_{2\bar{2}}\tan\delta_2 \\
   m_{2\bar{2}}\tan\delta_2 & 1-m_{\bar{2}\bar{2}}\tan\delta_2
   \end{array}\right|
  \\
 &=&
 {m_{02}\tan^2\delta_2\over m_{00}}
 \left|\begin{array}{cc}
   m_{02} &   m_{2\bar{2}} \\
   m_{0\bar{2}} & (\cot\delta_2-m_{\bar{2}\bar{2}})
   \end{array}\right|
 -{m_{0\bar{2}}\tan^2\delta_2\over m_{00}}
 \left|\begin{array}{cc}
   m_{02} & (\cot\delta_2-m_{22})\\
   m_{0\bar{2}} &   m_{2\bar{2}}
   \end{array}\right|
 \;.\nonumber
 \ea
 If the $d$-wave phase shift were
 small enough, it is easy to check that
 both Eq~(\ref{eq:result_D4}) and Eq~(\ref{eq:result_D2})
 simplifies to:
 \be
 \label{eq:simplify}
 \cot\delta_0(k)=m_{00}={\calZ_{00}(1,q^2;\eta_1,\eta_2)
 \over \pi^{3/2}\eta_1\eta_2 q}\;.
 \ee
 For the general case,
 Eq.~(\ref{eq:result_D4}) and Eq.~(\ref{eq:result_D2}) offer
 the desired relation between the energy eigenvalues in
 the $A^+_1$ sector and the scattering phases for the cases
 $\eta_1=\eta_2$ and $\eta_1\neq\eta_2$, respectively.

 \section{Large volume expansion of the scattering length}
 \label{sec:length}

 It is known that in low-energy scattering processes,
 the scattering phases $\delta_l(k)$ behaves
 like: $\tan\delta_l(k)\sim k^{2l+1}$ for small $k$,
 where $k$ is the relative momentum of the two particles being scattered.
 It is easy to verify that both $m_{00}$ and $m_{02}$
 behave like $1/q^3$ as $q\sim 0$.
 Since $\tan\delta_2(q)$ goes to zero like $q^5$,
 we see that the effects due to $d$-wave phase shift
 in Eq.~(\ref{eq:result_D4}) and  Eq.~(\ref{eq:result_D2})
 are negligible as long as the relative momentum $q$ is small enough.
 Therefore, in both cases, the $s$-wave scattering length $a_0$ will be
 determined by the zero momentum limit of Eq.~(\ref{eq:simplify}).

 For a large box, a large $L$ expansion of the formulae can be deduced.
 Using Eq.~(\ref{eq:simplify}),
 we find that the $s$-wave scattering length $a_0$
 is related to the energy difference in a generic
 rectangular box via:
 \footnote{
 For low relative momenta, the $d$-wave scattering phase behaves
 like: $\tan\delta_2(q)\sim a_2k^5=a_2(2\pi /L)^5q^5$, with
 $a_2$ being the $d$-wave scattering length.
 If we treat the effects due to $\tan\delta_2$
 perturbatively, we see from Eq.~(\ref{eq:result_D4}) and
 Eq.~(\ref{eq:result_D2}) that,
 in Eq.~(\ref{eq:scattering_length}), functions $c_1$ and $c_2$
 receive contributions that are proportional to $(a_2/L^5)$, which
 is of higher order in $1/L$ for large $L$.}
 \be
 \label{eq:scattering_length}
 \delta E=-{2\pi  a_0\over \eta_1\eta_2\mu L^3}
 \left[
 1+c_1(\eta_1,\eta_2)\left({a_0\over L}\right)
  +c_2(\eta_1,\eta_2)\left({a_0\over L}\right)^2 +\cdots
 \right]\;.
 \ee
 Here, $\mu$ designates the reduced mass of the
 two particles whose mass values are $m_1$ and $m_2$,
 respectively. Energy shift $\delta E\equiv E-m_1-m_2$ where
 $E$ is the energy eigenvalue of the two-particle state.
 Functions $c_1(\eta_1,\eta_2)$ and $c_2(\eta_1,\eta_2)$ are
 given by:
 \ba
 c_1(\eta_1,\eta_2)&=&{\hat{Z}_{00}(1,0;\eta_1,\eta_2)
 \over\pi\eta_1\eta_2} \;,
 \nonumber \\
 c_2(\eta_1,\eta_2)&=&{\hat{Z}^2_{00}(1,0;\eta_1,\eta_2)-
 \hat{Z}_{00}(2,0;\eta_1,\eta_2)\over
 (\pi\eta_1\eta_2)^2}\;,
 \ea
 where the subtracted zeta function is defined as:
 \be
 \label{eq:sub_zeta_def}
 \hat{Z}_{00}(s,q^2;\eta_1,\eta_2)=
 \sum_{|\tilde{\bn}|^2\neq q^2} {1 \over (\tilde{\bn}^2-q^2)^s}\;.
 \ee

 \begin{table}[htb]
 \caption{Numerical values for the subtracted zeta functions
 and the coefficients $c_1(\eta_1,\eta_2)$ and $c_2(\eta_1,\eta_2)$
 under some typical topology. The three-dimensional rectangular
 box has a size $L_1=\eta_1L$, $L_2=\eta_2L$ and $L_3=L$.
 \label{tab:numerical_values}}
 \begin{center}
 \begin{tabular}{c|l|l|r|r|r|r}
 \hline
 $L_1:L_2:L_3$ & $\eta_1$ & $\eta_2$
 & $\hat{Z}_{00}(1,0;\eta_1,\eta_2)$ & $\hat{Z}_{00}(2,0;\eta_1,\eta_2)$%
 & $c_1(\eta_1,\eta_2)$ & $c_2(\eta_1,\eta_2)$ \\
 \hline
 $1:1:1$ & $1$    & $1$   & $-8.913633$  & $16.532316$ & $-2.837297$ & $6.375183$\\
 $6:5:4$ & $1.5$ & $1.25$ & $-12.964476$ & $41.526870$ & $-2.200918$ & $3.647224$\\
 $4:3:2$ & $2$  & $1.5$   & $-16.015122$ & $91.235227$ & $-1.699257$ & $1.860357$\\
 $3:2:2$ & $1.5$    & $1$ & $-10.974332$ & $32.259457$ & $-2.328826$ & $3.970732$\\
 $2:1:1$ & $2$    & $1$   & $-11.346631$ & $63.015304$ & $-1.805872$ & $1.664979$\\
 $3:3:2$ & $1.5$  & $1.5$ & $-14.430365$ & $53.784051$ & $-2.041479$ & $3.091200$\\
 $2:2:1$ & $2$  & $2$     & $-18.430516$ & $137.771800$ & $-1.466654$ & $1.278623$\\
 \hline
 \end{tabular}
 \end{center}
 \end{table}
 In Table~\ref{tab:numerical_values}, we have listed
 numerical values for the coefficients
 $c_1(\eta_1,\eta_2)$ and $c_2(\eta_1,\eta_2)$
 under some typical topology. In the first column of
 the table, we tabulated the ratio for the three sides
 of the box: $\eta_1:\eta_2:1$. Note that for $\eta_1=\eta_2=1$,
 these two functions reduce to the old numerical values
 for the cubic box which had been used in earlier
 scattering length calculations.

 \section{Conclusions}
 \label{sec:conclude}

 In this letter, we have studied two-particle scattering states
 in a generic rectangular box with periodic boundary conditions.
 The relations of the energy eigenvalues and the scattering phases
 in the continuum are found. These can be viewed as a generalization of
 the well-known L\"uscher's formula.
 In particular, we show that the $s$-wave scattering length is
 related to the energy shift by a simple formula, which is a
 direct generalization of the corresponding formula in the case of
 cubic box. We argued that this asymmetric topology
 might be useful in practice since it provides more available low-lying
 momentum modes in a finite box, which will be
 advantageous in the study of scattering phase shifts at
 non-zero three momenta in hadron-hadron scattering and
 possibly also in other applications.


\end{document}